\documentclass[sn-mathphys,Numbered]{sn-jnl}

\usepackage{graphicx}%
\usepackage{multirow}%
\usepackage{amsmath,amssymb,amsfonts}%
\usepackage{amsthm}%
\usepackage{mathrsfs}%
\usepackage[title]{appendix}%
\usepackage{xcolor}%
\usepackage{textcomp}%
\usepackage{manyfoot}%
\usepackage{booktabs}%
\usepackage{algorithm}%
\usepackage{algorithmicx}%
\usepackage{algpseudocode}%
\usepackage{listings}%

\usepackage{soul,bm}
\usepackage{epstopdf,hyperref}
\usepackage{braket,dsfont,nicefrac}
\usepackage[mathcal]{euscript}
\hypersetup{colorlinks=true,citecolor=blue,linkcolor=magenta}
\sethlcolor{yellow}
\allowdisplaybreaks

\usepackage{xcolor}

\newcommand{\kk}{\mathbf{k}}

\newcommand{\win}{{\omega_\text{in}}}

\theoremstyle{thmstyleone}%
\theoremstyle{thmstyletwo}%

\theoremstyle{thmstylethree}%

\begin{document}

\title[Auger spectroscopy beyond the ultra-short core-hole relaxation time approximation]{Auger spectroscopy beyond the ultra-short core-hole relaxation time approximation}

\author[1]{\fnm{Alberto} \sur{Nocera}}\email{alberto.nocera@ubc.ca}
\affil[1]{\orgdiv{Stewart Blusson Quantum Matter Institute}, \orgname{University of British Columbia}, \orgaddress{\city{Vancouver}, \state{British Columbia}, \postcode{ V6T 1Z4},\country{Canada}}}
\author[2]{\fnm{Adrian} \sur{Feiguin}}\email{a.feiguin@northeastern.edu}
\affil*[2]{\orgdiv{Department of Physics}, \orgname{Northeastern University},\orgaddress{ \city{Boston}, \state{Massachusetts}, \postcode{02115}, \country{USA}}}

\abstract{
We present a time-dependent computational approach to study Auger electron spectroscopy (AES) beyond the ultra-short core-hole relaxation 
time approximation and, as a test case, we apply it to the paradigmatic example of a one-dimensional Mott insulator represented by a half-filled Hubbard chain. The Auger spectrum is usually calculated by assuming that, after the creation of a core-hole, the system thermalizes almost instantaneously. This leads to a relatively simple analytical expression that uses the ground-state with a core-hole as a reference state and ignores all the transient dynamics related to the screening of the core-hole. In this picture, the response of the system can be associated to the pair spectral function. On the other hand, in our numerical calculations, the core hole is created by a light pulse, allowing one to study the transient dynamics of the system in terms of the pulse duration and in the non-perturbative regime. Time-dependent density matrix renormalization group calculations reveal that the relaxation process involves the creation of a polarization cloud of doublon excitations that have an effect similar to photo-doping. As a consequence, there is a leak of spectral weight to higher energies into what otherwise would be the Mott gap. For longer pulses, these excited states, mostly comprised of doublons, can dominate the spectrum. By changing the duration of the light-pulse, the entire screening process can be resolved in time.  
}

\maketitle

\section{Introduction}
Auger recombination is a non-radiative decay process in which a conduction or valence electron can annihilate a hole, while the excess energy is transferred to a third carrier that undergoes a transition to a higher energy state. 
In Auger electron spectroscopy (AES)\cite{weightman1982,weissmann1981} a hole in a deep core energy level recombines with a conduction or valence electron and the energy released by this process is used to eject a second electron into the continuum. As a consequence, the valence or conduction band has lost two electrons, leaving behind a pair of holes. The ejected electron can be captured by a detector and, depending on the energy of the incident X-ray, it has in general a smaller energy compared with the first emitted (or primary) electron. In AES, the core-hole is the product of X-ray photons kicking out a $K$ or $L$ electron from the inner shells of the atom (notably, Auger emission has also been detected from annihilating core-hole electrons with low-energy positrons\cite{Weiss1989,Mukherjee2010,Chirayath2017,Fairchild2022}). Since these are mostly localized states, the resulting measurement contains information about the local correlations between electrons on the same atom. While in semiconductors and weakly interacting metals the spectrum can be interpreted as the self-convolution of the density of states \cite{Lander1953}, in correlated materials and superconductors it contains information about the electronic interactions \cite{Sawatzky1977, Sawatzky1980, Boer1984, Cini1977, Cini1979, Gunnarsson1980}. For instance, in the prototypical case of the Hubbard model with local interaction $U$, bound states of two-holes appear as sharp lines that can contain most of the spectral weight \cite{Sawatzky1980, Rausch2016}, which have been observed in core valence-valence (CVV) Auger electron spectroscopy experiments on Ni \cite{hufner1975,Maartenson1980,Sinkovic1997,kamakura2004,Born2021}. 

Despite considerable progress, the competition of many energy and time scales makes a microscopic understanding of the AES spectrum in such correlated states of matter a formidable task even in thermal equilibrium. If one considers the recent progress in time-resolved atomic spectroscopies\cite{beye2013,lu2020}, one can envision that pump-probe-like Auger experiments might be possible in the near future, rendering theoretical efforts in this direction valuable~\cite{Rausch2019}.

\begin{figure}
\centering
\includegraphics[width=0.48 \textwidth]{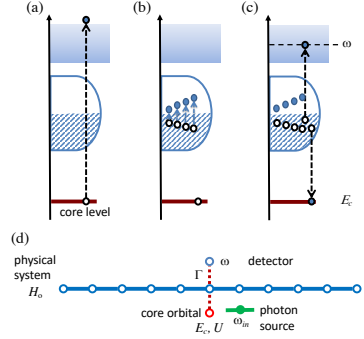}
\caption{Schematic representation of the regimes involved in the core-hole creation and non-radiative Auger recombination: (a) core-hole creation; (b) transient regime in which the core hole is screened creating a polarizarion cloud; (b) an electron recombines with the core-hole while a second one is ejected into the continuum. (d) Geometry used in the calculations. }
\label{fig:2s}
\end{figure}

The theory to describe the scattering cross section of CVV Auger spectroscopies was extensibly discussed primarily in the works of Gunnarsson and Schonhammer \cite{Gunnarsson1980,Gunnarsson1981,Gunnarsson1982}, who derived, using time-dependent perturbation theory, a closed-form expression of the AES scattering amplitude in terms of the ground-state energy $E_0$, the energy of the core-hole $E_c$, the intermediate energies of the system with a core-hole $E_n$, and the final energy $E_f$ with two holes in the valence band after recombination (no core-hole):
\begin{equation}
    I(\omega)=4\pi^2\sum_f \left|\sum_n \frac{\langle f, N-2| \Delta_{0}|n,N\rangle_c {\ }_c\langle n,N |d_0|0,N\rangle}{(\omega_{in}+E_0-E_n-E_c+i\eta)} \right|^2 \delta(\omega+E_f-E_0+E_c).
    \label{eq:1step}
\end{equation}
In this formula (from now on referred-to as ``one-step'' model, see Appendix~\ref{appendix} for a derivation), the operator $d_0$ creates a hole in the localized core orbital at position ``0'' by acting on the ground state of the system $|0\rangle$, and $\Delta_0=c_{0\downarrow}c_{0\uparrow}$ creates a localized hole pair in the valence band at the same position. Here, $\omega_{in}$ is the energy of the incoming photon that destroys the core hole, in a process that is equivalent to X-ray photoemission spectroscopy (XPS). The (real) phenomenological constant $\eta$ is introduced to account for the core-hole life-time. Notice that the intermediate sum over states $|n,N\rangle_c$ that contain a core-hole is squared and its evaluation requires a full knowledge of the eigenspectrum and matrix elements making this an insurmountable task except in small systems where exact diagonalization is possible. 

Sawatzky and Cini \cite{Sawatzky1977, Sawatzky1980, Boer1984, Cini1977, Cini1979} simplified the treatment by making extra assumptions: besides the sudden approximation, they consider the relaxation time of the core-hole to be the shortest time-scale in the problem (hence, dubbed ultra-short core-hole relaxation time approximation). In this context, the processes take place in two steps (that is why this is also referred-to as the ``two-step model''): first, the core hole is created and the system thermalizes, after which the Auger recombination takes place.  Therefore, only matrix elements between the initial state with a thermalized core-hole and the final state with two valence holes and no core-hole are involved. It is usually assumed that the system relaxes to the new ground state, resulting in an expression that resembles a two-particle spectral function:
\begin{equation}
I_{2s}(\omega)=\sum_f |\langle f,N-2|\Delta_0|0,N\rangle_{c}|^2 \delta(\omega + E_0' - E_f).
\label{eq:2step}
\end{equation}
Notice that the intermediate screening transient effects and the dependence with the core-hole potential have disappeared, being replaced by the wave function $|0,N\rangle_{c}$, representing the ground-state with the core-hole with energy $E_0'$. However, it is known that core-hole creation and recombination are largely non-equilibrium processes in which the creation of the core-hole shakes up the system inducing many particle-hole excitations in the valence band that appear as satellite peaks in the experiments \cite{Gunnarsson1980}
. Therefore, properly accounting for the finite core-hole lifetime and transient processes is crucial to describe the observed spectra. 

In this work, we introduce a time-dependent formulation of the problem that dispenses of perturbation theory and of several of the assumptions above and treats all energy and time scales on equal footing. Particularly, our approach accounts for effects arising from the finite life-time of the core-hole, transient processes, and finite width of the light pulse that creates the core-hole by explicitly solving the time-dependent Schr\"odinger equation and simulating the scattering event in real time. The paper is organized as follows: In sec.\ref{sec:methods} we introduce the model and computational details of the simulations; in sec.\ref{sec:results} we extensively describe results obtained by means of perturbative and non-perturbative approaches, establishing limits of validity, transient effects, and signatures of finite-core-hole relaxation. We finally close with a discussion.

\section{Model and methods} \label{sec:methods}

\subsection{The ``Auger Hamiltonian''}
We formulate the problem starting from ``first-principles'', including all the degrees of freedom involved in the scattering processes. Following the description in \cite{Gunnarsson1980, Cini1977, Rausch2019} we use a setup illustrated in Fig.\ref{fig:2s}(d). Since the process is mostly local, we ignore the momentum and lattice dependence and we consider the single-site approach. At the site of interest at position ``0'' we introduce a core-orbital and a source orbital for photons.
The problem is described by the Hamiltonian:
\begin{eqnarray}
H&=&H_0+H_c+\win b^\dagger b+\sum_{\kk,\sigma} \omega_\kk \tilde{c}^\dagger_{\kk\sigma}\tilde{c}_{\kk\sigma} + V_A+V_{c}, \nonumber \\
V_A&=& \Gamma\sum_{\kk,\sigma} \tilde{c}^\dagger_{\kk\sigma}d^\dagger_{0\sigma} \Delta_0  + {\mathrm h.c.} \nonumber \\
V_c&=& \sum_{\kk, \sigma} J'(\kk) \tilde{c}^\dag_{\kk\sigma} d_{0,\sigma}(b+b^\dagger)+{\mathrm h.c.} \nonumber \\
H_c&=&-U_c \sum_\sigma (1-n_{d0\sigma})n_{d0}+E_c n_{d0},
\label{eq:Htot1}
\end{eqnarray}
where $H_0$ is the lattice Hamiltonian, $\omega_{in}$ is the energy of the incoming photons, which will be resonantly tuned to the energy of the XPS emission edge as we describe below. The term $V_c$ annihilates a photon and creates a core-hole with $d_0$ and the Auger term $V_A$ annihilates the core-hole with a valence electron, effectively creating a hole pair in the valence band and promoting an electron to the scattering continuum through operator $\tilde{c}^\dagger_\kk$. In addition, upon the creation of the core-hole, electrons experience a localized attractive potential parametrized by $U_c$.   
The time scale for the Auger process is determined by $1/\Gamma$, while the one for the core-hole creation, by $1/J'$. 
In order to make the problem tractable we introduce a further simplification by ignoring matrix element effects, such as the dependence of $\Gamma$ and $J'$ on $\kk$. In addition, we neglect the effects of the \emph{first} photoemitted electron after the creation of the core-hole and we consider single photon processes, so only one photon states are included in the problem. Similar approaches have been used to study non-equilibrium Auger processes \cite{Rausch2019}, non-equilibrium X-ray absorption\cite{Werner2022}, resonant inelastic X-ray Scattering (RIXS)\cite{Eckstein2021,Zawadzki2020,Zawadzki2020b} and non-equilibrium photoemission\cite{Zawadzki2019,Cohen2014,kantian2015}.
According to these considerations, the Hamiltonian we use becomes:
\begin{eqnarray}
H&=&H_0+H_c+V_A+V_{c}+ \nonumber \\
&+&\win b^\dagger b+ {\omega_\text{photo}\sum_{\sigma}\tilde{c}'^\dagger_{\sigma}\tilde{c}'_{\sigma}} +\omega_\text{Auger}\tilde{c}^\dagger_\uparrow\tilde{c}_{\uparrow} \nonumber \\
V_A&=& \Gamma \sum_\sigma\tilde{c}^\dagger_{\sigma}d^\dagger_{0\sigma} \Delta_0  + {\mathrm h.c.} \nonumber \\
V_c&=& J' \tilde{c}'^\dagger_\uparrow d_{0,\uparrow}(b+b^\dagger)+{\mathrm h.c.} \nonumber \\
H_c&=&-U_c (1-n_{d0\uparrow})n_{d0}+E_c n_{d0}.
\label{eq:Htot}
\end{eqnarray}
Here, we have included only two scattering states corresponding to the photoexcited and Auger electrons, $\tilde{c}'$ and $\tilde{c}$, respectively. 
This corresponds to having a ``detector'' tuned to a particular scattering energy $\omega_{\text{Auger}}$ and we are assuming that these two electrons are distinguishable and independent. If initially the core orbital is double occupied and we have a photon at energy $\win$, after photoexcitation via the $V_c$ term, the photon and an $\uparrow$ core electron with energy $E_c$ are annihilated, and a photelectron with energy $\omega_\text{photo}$ is created. At the same time, the Auger electron with energy $\omega_{\text{Auger}}$ is emitted by the $V_A$ term, so that the overall energy conservation for a zero temperature equilibrium situation reads: $\win+E^{N}_0 =\omega_{\text{photo}}+\omega_{\text{Auger}}+E^{N-2}_{\beta}$, where $E^{N-2}_{\beta}$ is a generic excited state energy of the system with two missing electrons in the valence band after the Auger electron emission.
Notice that there is freedom on the choice of $\omega_{\text{photo}}$ which is discussed in the next section. In fact in the dDMRG calculations we will pick $\omega_{\text{photo}}\approx\omega_{\text{edge}}$, from the main dominant peak of the XPS spectrum.  

\subsection{One-step model}
Following the seminal work of Gunnarsson and Sch\"{o}nhammer\cite{Gunnarsson1980}, the one-step 
formula for the Auger electron emission can we written as (see Appendix~\ref{appendix} for details):

\begin{align}
I_{\text{Auger}}(\omega)&=-\frac{1}{\pi}\textrm{Im}\Big[\langle\phi_{e}|\Delta^\dag_{0}\frac{1}{\omega-H_0+E_0+i\eta}\Delta_{0} |\phi_e\rangle\Big].
\label{eq:1stepb}
\end{align}
where, from the conservation of energy, we have that $\omega=\omega_{\text{Auger}}-\omega_{\text{photo}}$.
 In the derivation of this expression we assume that the core hole is created at the XPS ``emission edge''.
 As previously anticipated, in this work we neglect matrix elements effects and their momentum dependence, as well as 
the effect of Coulomb interactions 
between core electrons. 
Therefore, 
 the state $|\phi_e\rangle$ is defined as:

\begin{equation}\label{Auger_CV}
 |\phi_e\rangle = \frac{1}{\omega_{\text{edge}}-H_0-H_c+E_0+i\eta} |N,0\rangle.
\end{equation}

In order to determine the emission edge $\omega_{\text{photo}}\approx\omega_{\text{edge}}$ we first need to obtain the XPS or X-ray photoemission spectrum. 
From Ref.~\cite{Gunnarsson1980}, the XPS spectrum (ignoring
again matrix element effects and assuming as usual the sudden approximation to be valid) is given by

    
\begin{equation}
I_{XPS}(\omega_{\text{photo}}) = -\frac{1}{\pi}\textrm{Im}\Big[\langle N,0|\frac{1}{\omega_{\text{photo}}-H_0-H_c+E_0+i\eta}|N,0\rangle\Big],
\label{eq:1stepc}
\end{equation}
where, from the conservation of energy, we have that $\omega_{\text{photo}}=E_c+\win$.
Expressions (\ref{eq:1stepb}) and (\ref{eq:1stepc}) are computed using the dynamical density matrix renormalization group method (dDMRG)\cite{kuhner1999,jeckelmann2002,nocera2016} (see Results section with more details about how these calculations are performed).

\begin{figure}
\centering
\includegraphics[width=0.48 \textwidth]{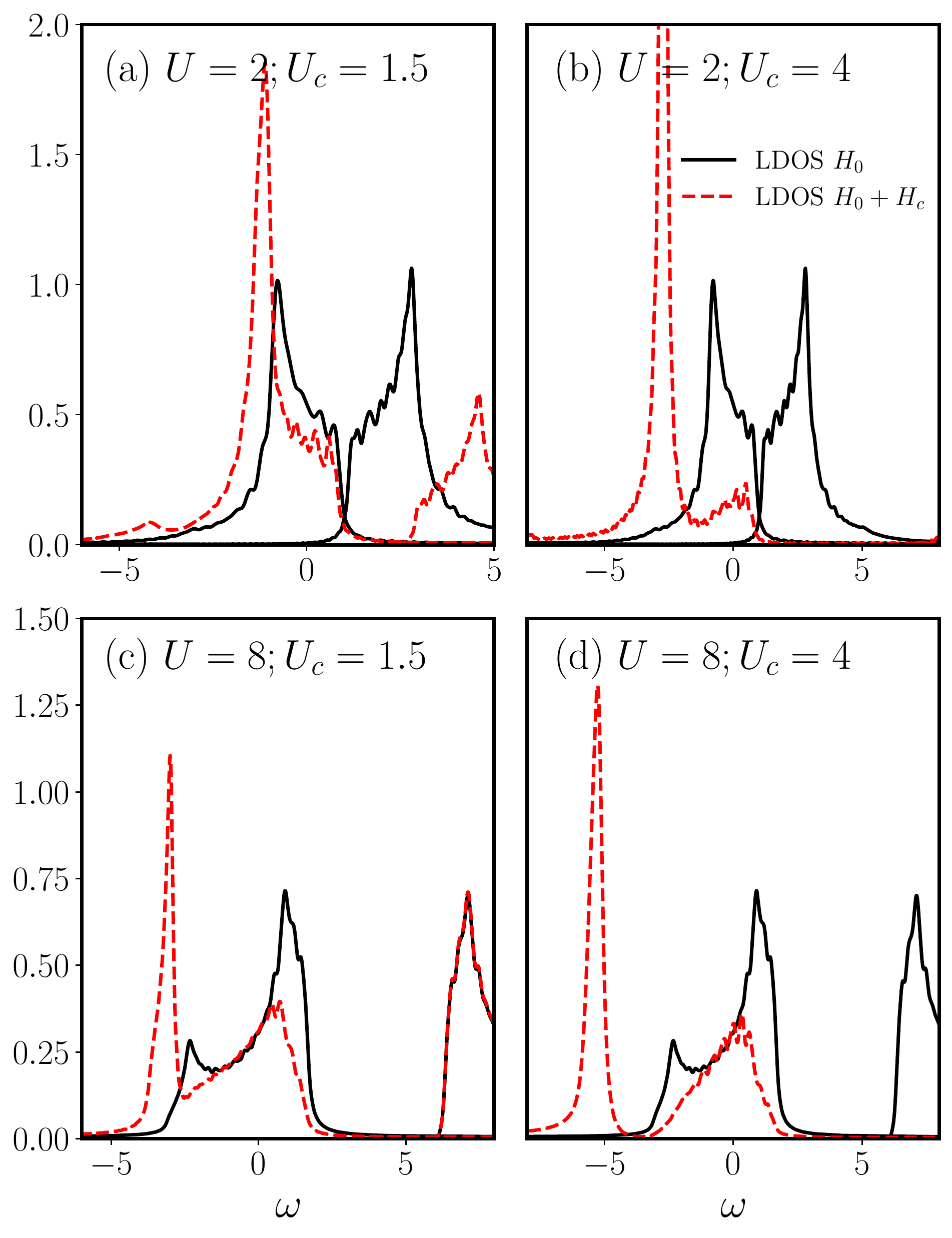}
\caption{Local density of states for the system with and without a core-hole for several values of $U$ and $U_c$. There is no time dependence in the spectrum since we assume that the initial states with the core-hole has thermalized. For large values of $U_c$, the core hole induces the formation of a localized bound  state. }
\label{fig:ldos}
\end{figure}

\subsection{Two-step model}

In all the cases studied below, we compare to results obtained using the two step formula, Eq.~(\ref{eq:2step}). 
Notice, by looking at Eqs.~(\ref{eq:1stepb}) and (\ref{eq:2step}), that the main difference between both expressions is on the definition of the initial state used in the spectral function.
Both quantities can be obtained by using either the dDMRG or tDMRG methods. The tDMRG calculation consists of evaluating a modified two-particle spectral function\cite{Rausch2016} in real time 
$$G_{2s}(t)={}_{c}\langle 0,N|\Delta^\dagger_0(t) \Delta_0(t=0) |0,N\rangle_c,$$
where $\Delta^\dagger_0(t)=\exp{(iH_0t)}\Delta^\dagger_0\exp{(-iH_0t)}$.
Here, $|0,N\rangle_c$ is the ground state of Hamiltonian $H=H_0+H_c$ with the core-hole. After annihilating a doublon with $\Delta_0$ at the reference site ``0'' (at the center of the chain) the system is time-evolved with the Hamiltonian without the core-hole $H_0$. A Fourier transform with an artificial Lorentizan broadening of $\epsilon=0.1$ yields the frequency resolved response $I_{2s}(\omega)=-1/\pi {\rm Im}G_{2s}(\omega)$\cite{white2004a,Paeckel2019}. 

\subsection{Time-dependent scattering approach}

The time-dependent scattering approach consists of directly simulating the evolution of the system under the action of Hamiltonian (\ref{eq:Htot}). In practice, it proceeds  as follows: at time $t=0$ the photon level is occupied by a single photon and the core orbital is double occupied. The lattice electrons may, or may not be in the ground state, since our formulation is generic and does not rely on any equilibrium assumptions. Then, the $V_c$ term is turned on for a finite time $t_{probe}$ and turned off, equivalent to a square pulse (we will not discuss the dependence of the results on the shape of the pulse), with $\win=\omega_{\text{edge}}$. The Auger term acts during the full time of the simulation. The occupation $\tilde{n}(t)$ of the scattering state is measured as a function of time during the duration of the experiment. As mentioned above, for practical purposes we only consider one scattering state at a time and we conduct independent simulations in parallel with varying $\omega\equiv\omega_{\text{Auger}}-\omega_{\text{photo}}$ . 
The time-dependent Schr\"odinger equation for the full problem involving all the degrees of freedom (incoming photon, lattice electrons, scattering state) is solved using the time-dependent density matrix renormalization group method (tDMRG)\cite{white2004a,daley2004,vietri,Paeckel2019}. Notice, however, that this approach does not depend on the Hamiltonian $H_0$, the geometry of the problem, or the solver used for the time evolution.

\begin{figure}
\centering
\includegraphics[width=0.48 \textwidth]{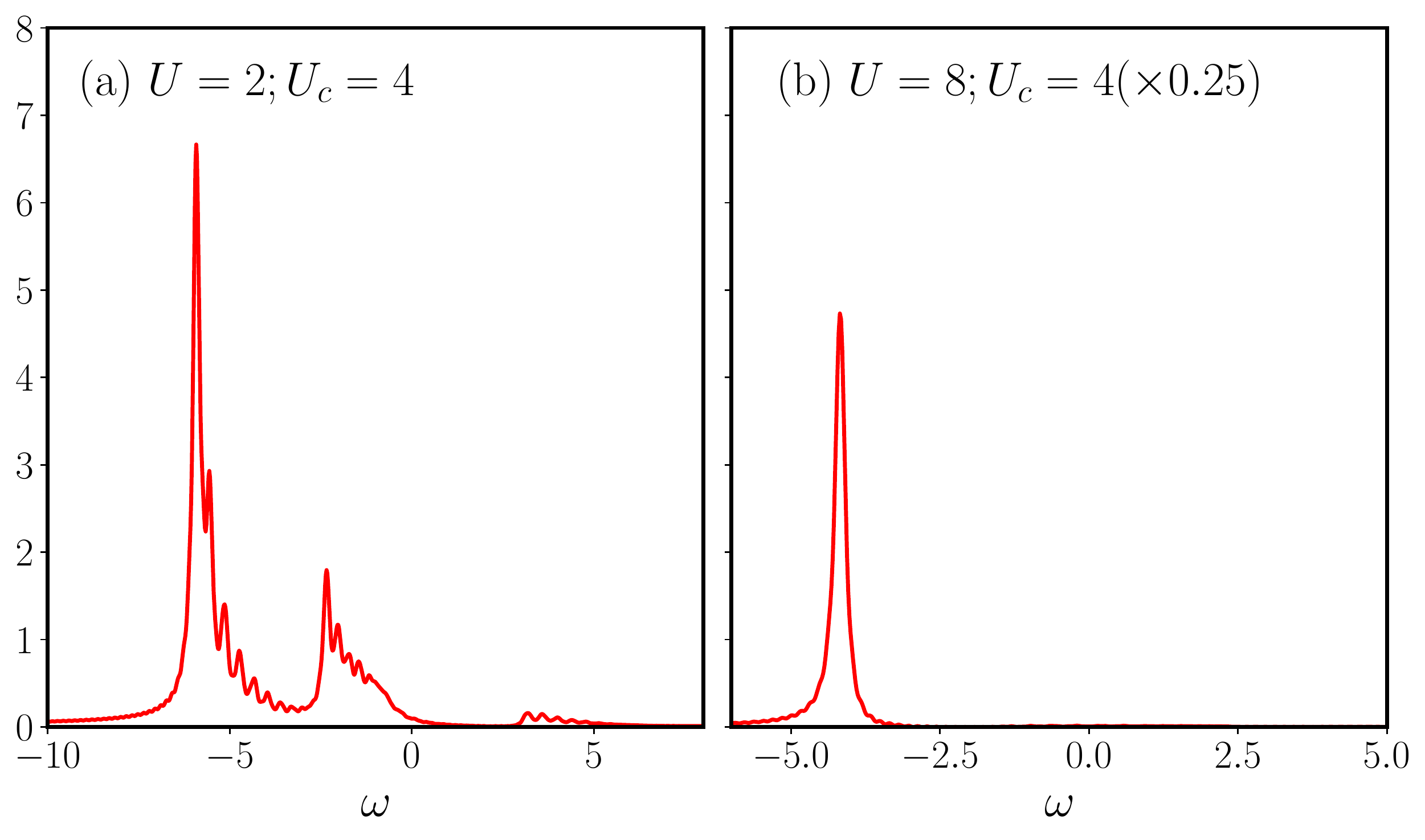}
\caption{X-ray photoemission spectra (XPS) as a function of $\omega\equiv\omega_{\text{photo}}$ for two values of $U=2,8$ and $U_c=4$. The energy of the incoming photon $\omega_{\text{photo}}$ is tuned to the position of the first peak. For the dDMRG simulations, we have used a broadening $\eta=0.1J.$}
\label{fig:xps}
\end{figure}


\section{Results}\label{sec:results}

As a proof of concept demonstration, we focus our attention on the paradigmatic Hubbard chain, described by the Hamiltonian:
\begin{equation}
H_0 = -J \sum_{i=1,\sigma}^{L-1} \left(c^\dagger_{i\sigma} c_{i+1\sigma}+\mathrm{h.c.}\right) + U \sum_{i=1}^L n_{i\uparrow}n_{i\downarrow}
\label{Hubbard}
\end{equation}
where the operator $c^\dagger_{i\sigma}$ creates an electron of spin $\sigma=\{\uparrow,\downarrow\}$ on the
$i^{\rm th}$ site along a chain of length $L$, $n_\sigma=c^\dagger_\sigma c_\sigma$, and the on-site Coulomb repulsion is parametrized by $U$. In the following, we set the electronic density to half-filling $N=L$ and consider several values of $U>0$. Our unit of energy is the hopping $J$ (we reserve the label $t$ to refer to ``time'') and time is measured in units of $1/J$. Unless otherwise stated, we consider a chain with $L=32$ sites with open boundary conditions, 200 DMRG states and a time step $\delta t=0.02$ with the value of the core-hole energy fixed to $E_c=-10J$.

We here provide details of our dDMRG calculations.
In order to compute the one-step formula in Eq.~(\ref{eq:1stepb}) using dDMRG, 
we first compute $|\phi_{e}\rangle$ in Eq.~(\ref{Auger_CV}) at incident energy $\omega_{edge}$  
with the standard Correction-Vector algorithm using the Krylov approach developed in \cite{nocera2016}. Following a similar procedure introduced for RIXS dDMRG calculations~\cite{nocera2018rixs},
we then the apply the operator $\Delta_0$ to $|\phi_{e}\rangle$ and compute 
for each energy $\omega$ 
the following correction-vector
\begin{equation}
    |x(\omega)\rangle = \frac{1}{\omega-H_0+E_0+i\eta}\Delta_0|\phi_e\rangle.
\end{equation}
The one-step formula in Eq.~(\ref{eq:1stepb}) is then readily evaluated as $I_{Auger}(\omega) = -\frac{1}{\pi} \text{Im}[\langle\phi_e|\Delta_0^\dagger|x(\omega)\rangle]$ for each energy $\omega$.
In our dDMRG calculations, we have used up to 1000 DMRG states making sure the truncation error is below $10^{-6}$. Finally, we have fixed a broadening $\eta=0.1J$.

\begin{figure}
\centering
\includegraphics[width=0.48 \textwidth]{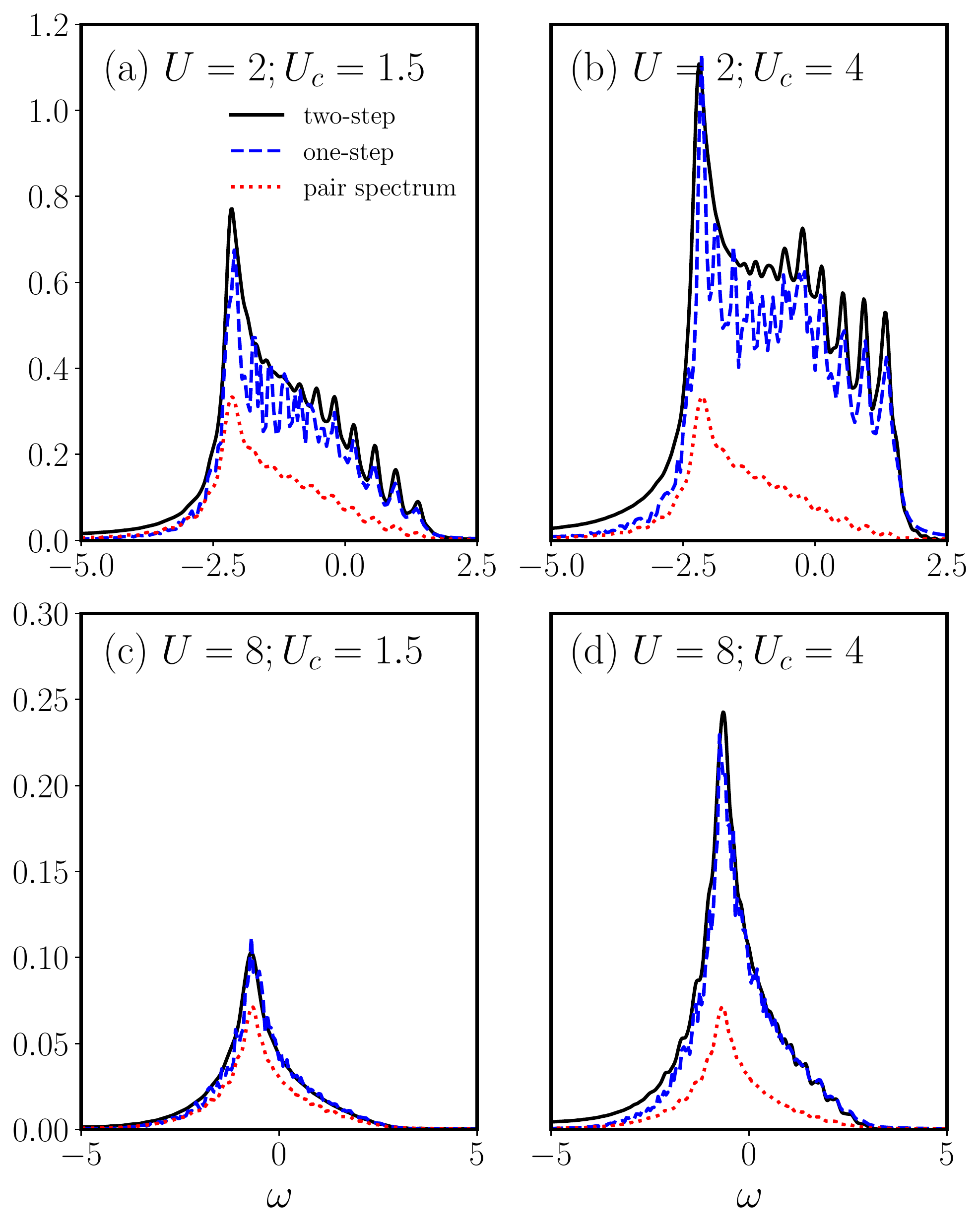}
\caption{AES spectrum as a function of $\omega\equiv\omega_{\text{Auger}}-\omega_{\text{photo}}$ calculated using the two-step model in Eq.~\ref{eq:2step} and the one-step model in Eq.~\ref{eq:1step} obtained with dDMRG ($\eta=0.1J$), compared to the pair spectrum.}
\label{fig:ddmrg}
\end{figure}

\subsection{Ultra-short relaxation}

In the $U=0$ limit and ignoring the effects of the core-hole potential, the Auger spectrum can be interpreted by generalizing Lander's intuitive picture\cite{Lander1953} to the case of partial density $n=N/L < 2$. In this regime, all electrons (and Auger pairs) are completely uncorrelated, and the spectrum is a self-convolution of the {\it occupied part} of the local density of states\cite{Potthoff1993,Rausch2016}:
\[
I_{2s}(\omega) \sim \int d\epsilon \rho(\omega)\rho(\omega-\epsilon).
\]
The energies $\epsilon$ allowed by energy conservation range from $E_c+2(\omega_0-E_c)$ to $E_c+2(\mu-E_c)$, where $\omega_0$ is the bottom of the valence band and $\mu$ is the chemical potential in equilibrium. Therefore, the bandwidth is roughly $2(\mu-\omega_0)$, twice the bandwidth of the photoemission spectrum. 

It is important to recognize that in a partially filled band and, more precisely, in a Mott insulator Lander's picture does not longer apply. This interpretation is strictly valid for the case of a fully filled band where the state $|0,N\rangle_c$ is an eigenstate of the system with and without the core-hole. However, at partial filling this is not the case: the initial state is not an eigenstate of $H_0$, but of $H_0+H_c$. Hence, the localized state has a finite overlap with all the states in the valence band. %

The role of the interaction $U$ and the core-hole potential cannot be underestimated: the Coulomb repulsion prevents double occupancy and, as $U$ increases, the system more closely resembles a spin chain. In this regime, Auger processes are greatly suppressed. At the same time, the influence of the local attractive core-hole potential becomes more relevant, since it counteracts the repulsion and tends to favor the formation of a localized state. To guide intuition, in Fig.~\ref{fig:ldos} we show the local density of states (LDOS) at site ``0'' for Hamiltonians $H_0+H_c$ and $H_0$ (with, and without the core-hole, respectively). In the presence of a core-hole potential we can clearly identify a sharp peak in the LDOS indicating localized bound states that will be responsible for the redistribution of the Auger spectral weight. 

The energy of the bound state determines the position of the X-ray emission edge, which is obtain from the position of the lowest peak in the XPS spectrum, as shown in Fig. ~\ref{fig:xps}. The numerical results for $\omega_\mathrm{edge}$ are listed in Table \ref{table:wedge}.

\begin{table}

\centering
\caption{Values of $\omega_\mathrm{edge}$ obtained through XPS calculations.}
\begin{tabular}{ c|c|c } 
$U$ & $U_c$ & $\omega_\mathrm{edge}$ \\
\hline
2 & 1.5 & -1.8 \\ 
2 & 4 & -5.9 \\ 
8 & 1.5 & -1.5 \\ 
8 & 4 & -2.4 \\

\end{tabular}
\label{table:wedge}

\end{table}

In Fig. ~\ref{fig:ddmrg} we show the Auger spectrum obtained with the two-step formula Eq.~\ref{eq:2step} and the one-step formula Eq.~\ref{eq:1step} evaluated as described in the previous section, together with the ``pair spectrum'' corresponding to the bare Hamiltonian with $U_c=0$. 
For small $U=2$, despite the spectral weight being different, the Auger spectrum and pair spectrum with and without the core-hole potential $U_c$ do not differ much, except for larger $U_c=4$. Therefore, it is tempting to identify the Auger spectrum with the pair spectrum. We will see that in our time dependent approach, this naive identification will no longer apply, and the spectrum will largely depend on the relaxation time scale of the core-hole.



\begin{figure}
\centering
\includegraphics[width=0.48 \textwidth]{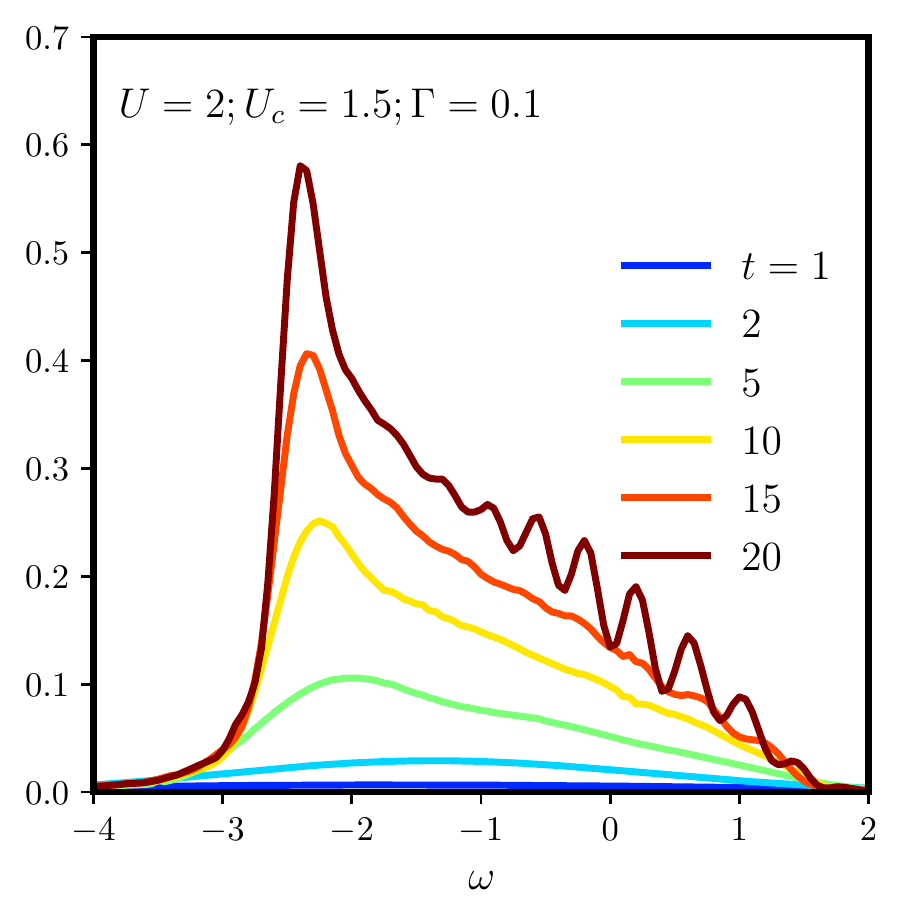}
\caption{Non-perturbative ``two-step'' calculation: Evolution of the spectrum as a function of time for $U=2; U_c=1.5$ and $\Gamma=0.1$. As time evolves, features appear sharper and resolution improves, as expected from the time-energy uncertainty principle. The spectra are plotted as a function of $\omega\equiv\omega_{\text{Auger}}-\omega_{\text{photo}}$}.
\label{fig:time}
\end{figure}

\begin{figure}
\centering
\includegraphics[width=0.48 \textwidth]{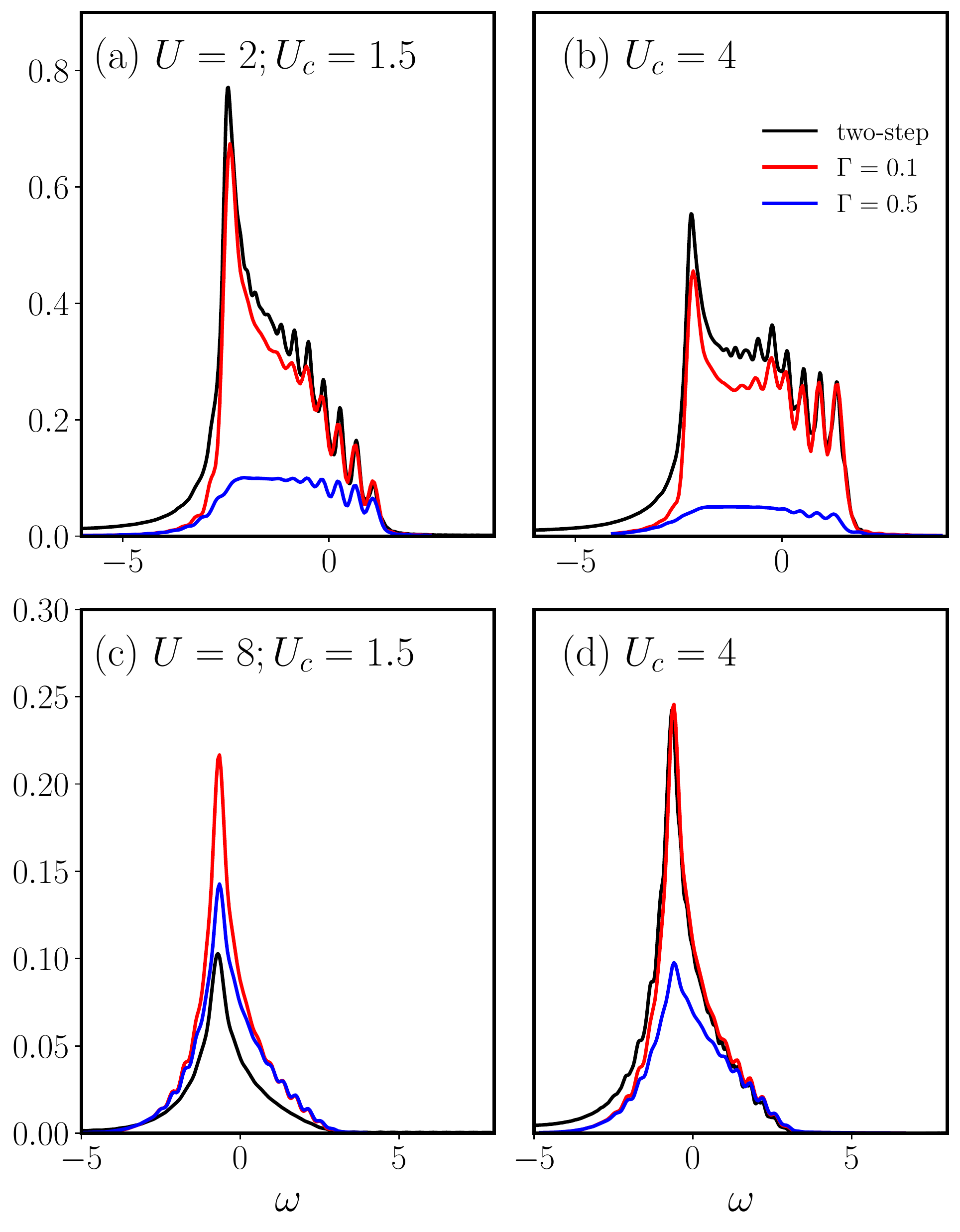}
\caption{Time dependent ($t=20$) non-perturbative Auger spectrum assuming a two-step model in which the initial state with the core-hole has thermalized. We show results for different values of $U; U_c$ and the Auger coupling $\Gamma=0.1,0.5$. Results have been rescaled to offer a better comparison. The spectra are plotted as a function of $\omega\equiv\omega_{\text{Auger}}-\omega_{\text{photo}}$}.
\label{fig:2step}
\end{figure}

\subsection{Non-perturbative effects}

In order to gain intuition and understand the limits of the theory, we first consider the case of a finite Auger coupling $\Gamma$ within the ultra short relaxation time approximation. This corresponds to using the ground state with a core-hole as our initial state. At time $t=0$ we start evolving the system with a single $\omega$ mode and we measure its population $\tilde{n}(t)$ as a function of time. Therefore, $V_c$ and the photon do not enter into the problem: we assume that this term acted earlier during the ``first step'' of the two-step model and subsequently the system equilibrated very fast. We will refer to this simulation as a ``non-perturbative'' two-step calculation.

As illustration, we show the evolution of the Auger spectrum in  Fig.~\ref{fig:time} for different values of Hubbard $U$, core-hole potential $U_c$ and Auger coupling $\Gamma$. As time evolves, the spectral features appear sharper and better defined, as expected from the energy-time uncertainty principle. 
In the following, in order to present the spectra on the same scale, we plot $\tilde{n}/\Gamma^2$ at time $t=20$. 


At small $\Gamma$, perturbation theory applies and we expect agreement with the two-step formula Eq.~(\ref{eq:2step}). This is confirmed in our calculations for $\Gamma=0.1$, as seen in Fig.~\ref{fig:2step}. We point out that the results have been rescaled to offer a better comparison (a scale factor has no physical significance, and even experimental data is usually normalized and usually displayed in ``arbitrary units''). This is telling us that for relatively small $\Gamma=0.1$ the Auger term indeed acts as a perturbation. 
Upon increasing the value of $\Gamma$ we start to observe a noticeable departure from the perturbative results, with a much more reduced weight and a modified lineshape. We believe that these effects are due to the formation of an entangled state between the conduction electrons, the core state at energy $E_c$ and the scattering electron at $\omega$. These results validate our time-dependent approach to the problem and illustrate how it is possible to resolve non-perturbative effects by solving the problem in the time domain instead of frequency. 

\begin{figure}
\centering
\includegraphics[width=0.48 \textwidth]{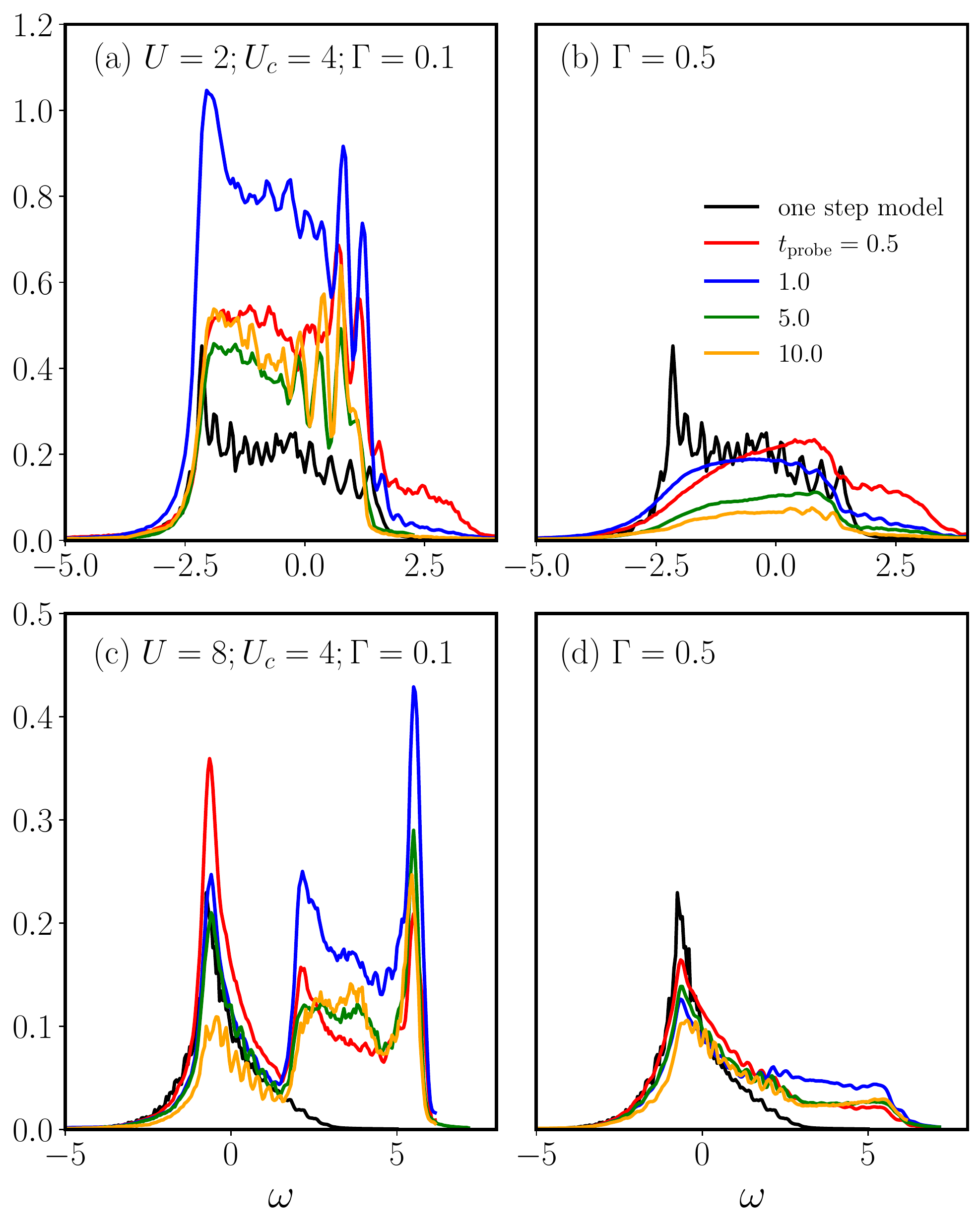}
\caption{Full time-dependent simulation of the Auger spectrum including the core-hole dynamics and transient. The pulse that creates a core hole has a duration $t_{probe}$ and the results are obtained at a time $t=20$ after the pulse. We focus on the case $U_c=4$. The spectra are plotted as a function of $\omega\equiv\omega_{\text{Auger}}-\omega_{\text{photo}}$}.
\label{fig:full}
\end{figure}

\subsection{Finite core-hole relaxation time}

In order to study the effects of the finite core-hole relaxation time, we carry out a full time-dependent simulation of the scattering event, including the scattering photons and accounting for transients during the creation of the core-hole. To this aim, we turn on the $V_c$ term for a time $t_{probe}$, during which a core electron interacts with the photon field. At the end of this time interval, the wave function will be in a superposition of the system with the core-hole and without. Since the states without the core-hole do not participate in the Auger dynamics, we project them out in order to gain numerical accuracy by applying the operation $|\tilde\psi\rangle=(1-n_{d\uparrow})|\psi(t)\rangle$. After the projection, the state is normalized again and the time-evolution resumed but without the $V_c$ term. Auger recombination occurs during the entire time of the simulation, including the duration of the light pulse,  transient regime, and the aftermath. At the start of the measurement, the system is far from equilibrium, a situation depicted in Fig.~\ref{fig:2s}(b). In this case, one can neither define a chemical potential, nor a temperature for the electrons. 

\begin{figure}
\centering
\includegraphics[width=0.48 \textwidth]{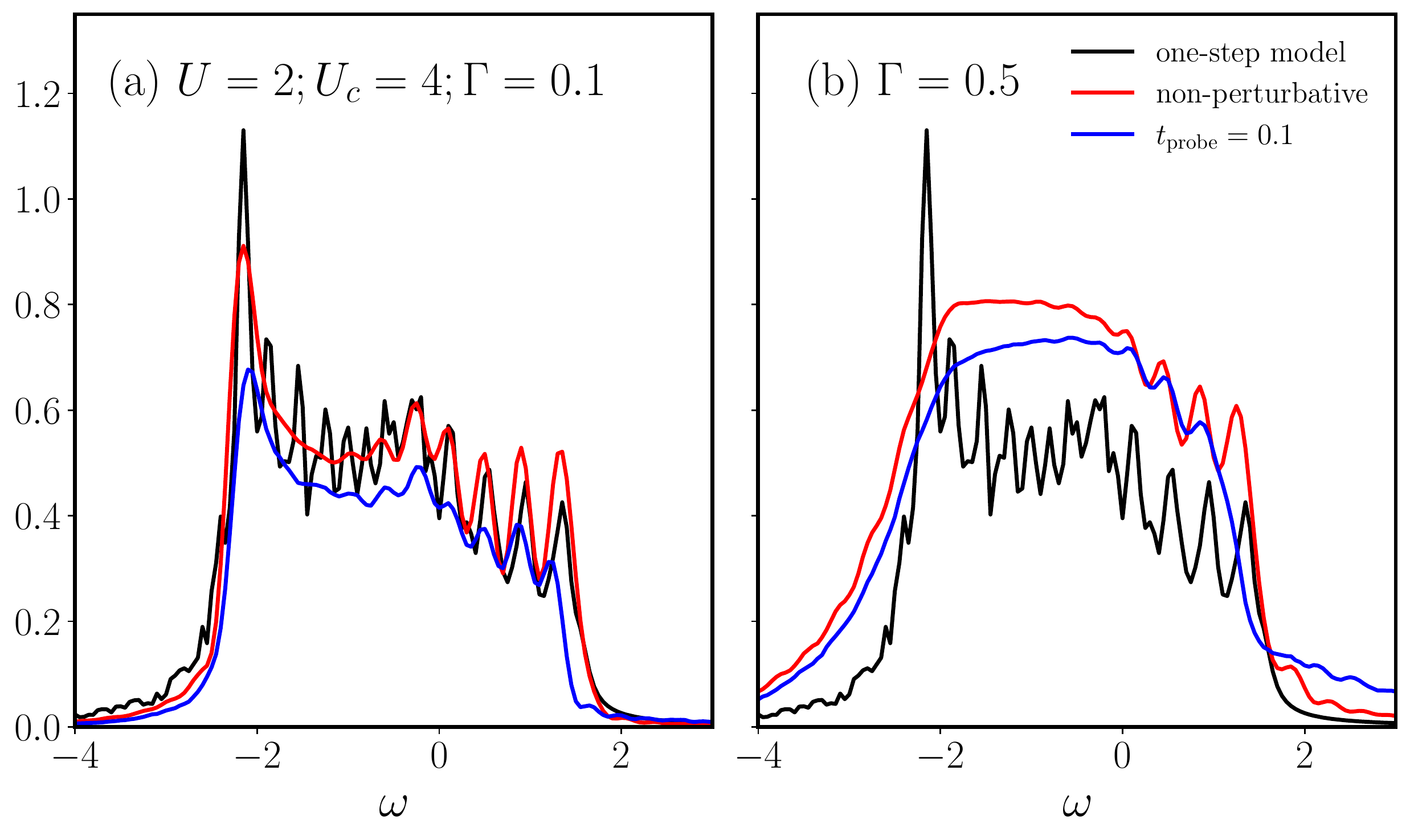}
\caption{Comparison of three approaches discussed in this work:  perturbative one-step model using DDMRG, non-perturbative two-step model, and full simulation with $t_{probe}=0.5$ for a choice of parameters $U=2$, $U_c=4$ and $\Gamma=0.1, 0.5$. The non-perturbative results were rescaled by a factor 2 in panel (a) and a factor 4 in panel (b). The spectra are plotted as a function of $\omega\equiv\omega_{\text{Auger}}-\omega_{\text{photo}}$}.
\label{fig:comparison}
\end{figure}

In Fig.~\ref{fig:full} we show results for several values of $U$, $\Gamma=0.1,0.5$ and $U_c=4$. At short time $t_{probe}$, the system does not have time to equilibrate after the core-hole is created, and the response resembles the results for the one-step formula, particularly for small $\Gamma=0.1$. On the other hand, for large $t_{probe}$ we observe a considerable transfer of spectral weight to higher energies, both for large and small $\Gamma$. This indicates that we are observing a true non-equilibrium effect originating from the finite duration of the pump and the long relaxation time of the hole.

This leads to the obvious question: why there is so much spectral weight transferred to higher energies?
Notice that after the core-hole is created the system does never relax to the ground state because the chain is finite and there are no dissipation mechanisms. This implies the presence of a large number of particle-hole excitations. 
In a Mott insulator, the ``particle'' excitations are doublons in the upper Hubbard band. Therefore, there is no a priory reason to assume that the system is close to the initial state of the two-step approximation. One would therefore expect a noticeable departure between the results obtained with the one or two-step models and the full time-dependent calculation. In Fig.~\ref{fig:comparison} we show a comparison of the three approaches (perturbative one-step model of Eq.~(\ref{eq:1step}), non-perturbative two-step model and full simulation) for a particular choice of parameters as illustration. We have multiplied the non-perturbative results by a scale factor in order to make the similarities more evident. We see that for small $\Gamma=0.1$ all approaches agree quite well, but for $\Gamma=0.5$ the non-perturbative approaches agree with each-other (as long as $t_{probe}$ is kept small) and differ form the perturbative results. This confirms the validity of the two-step model as long as $\Gamma$ and the pulse duration are kept small. 

\begin{figure}
\centering
\includegraphics[width=0.48 \textwidth]{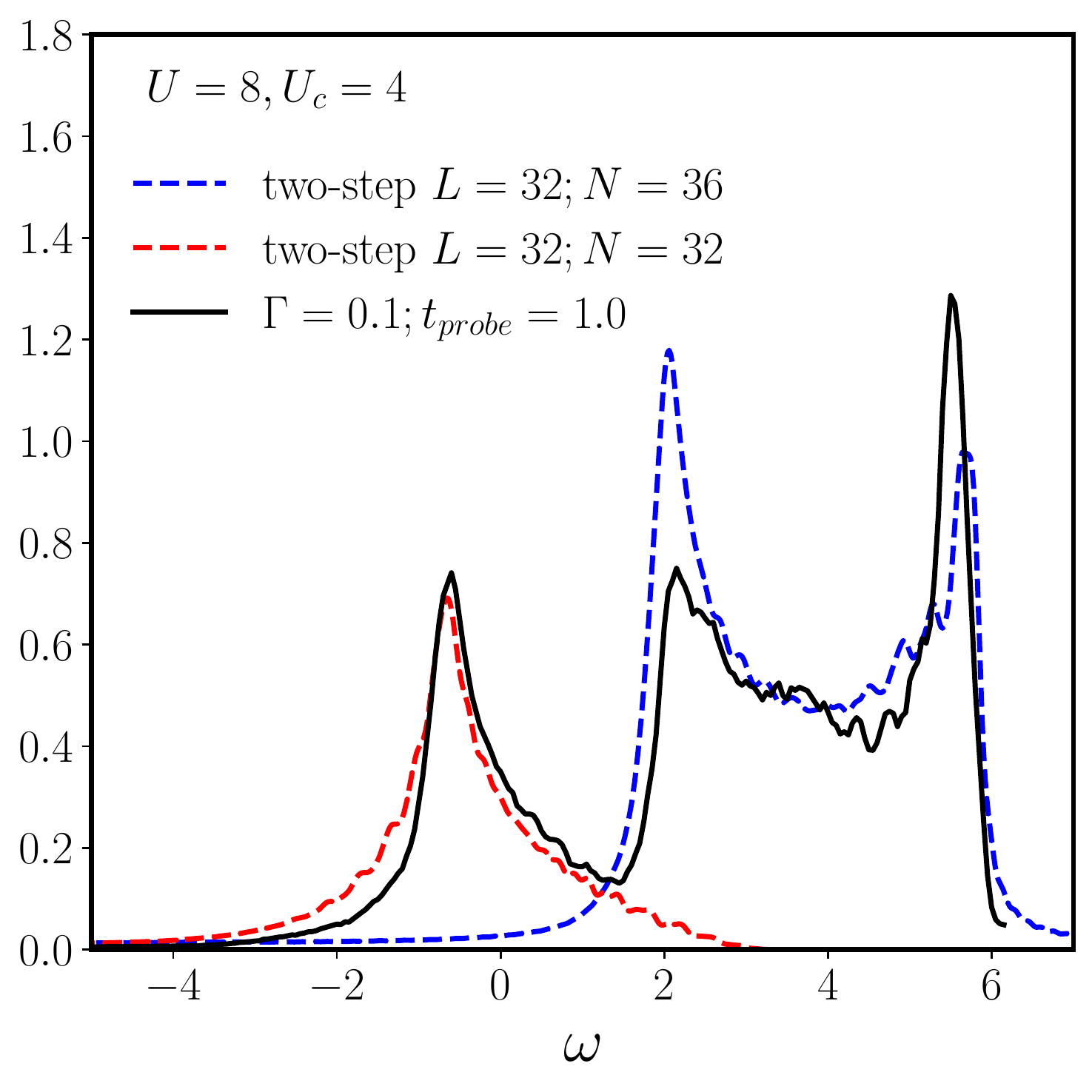}
\caption{Full simulation compared to the two-step approach with density $n=36/32$, $U=8$, $U_c=4$ and $\Gamma=0.1$. These results demonstrate that the high energy spectral weight originates from a ``photo-doping''-like effect: The core-hole forms a screening cloud of doublon excitations that are responsible for most of the high energy features. The spectra are plotted as a function of $\omega\equiv\omega_{\text{Auger}}-\omega_{\text{photo}}$}. 
\label{fig:doped}
\end{figure}

Beyond this regime of validity, the screening effects are more dramatic, particularly for larger $U_c=4$ (see Fig.~\ref{fig:full}(c), for instance). The high energy features stem from a cloud of doublons screening the core-hole and allow one to resolve the evolution of the entire screening process in time. To support this argument, we show results using the two-step formula for a density $n>1$ in Fig.~\ref{fig:doped}. We observe that the extra electrons that concentrate around the core-hole are responsible for the spectral weight. Notice that, even though doublons reside in the upper Hubbard band above the Mott gap, once the upper Hubbard band is populated, the Auger spectrum becomes gapless.  

\section{Summary and Conclusions}

The Auger spectrum is usually interpreted within a two-step approach as the pair spectral function. This assumes that core-holes thermalize very fast after being created and ignores the transient dynamics that takes place due to the presence of the attractive core-hole potential. In reality, core-holes are screened by an electronic cloud in a process that is very far from equilibrium and requires for one to be able to account for higher order contributions in a perturbative sense. Moreover, in the absence of dissipation mechanisms, the system does not thermalize to the ground state. The one-step formula Eq.~(\ref{eq:1step}) includes important corrections but it is computationally challenging to evaluate due to the presence of matrix elements that involve all energy eigenstates. 

Our time-dependent scattering approach is formulated in the time-domain instead of the frequency domain and allows one to carry out a ``numerical experiment'' that includes all the degrees of freedom involved in the scattering event, namely photons, core and conduction/valence electrons, and scattering states. In a nutshell, it is equivalent to sending a light pulse of duration $t_{probe}$ to create a core-hole, and measuring the response of the system at the detector which captures the emitted Auger electron at a given energy {\it in real time}. Since the light-matter coupling is the smallest energy scale in the problem, it suffices with keeping states that involve a single photon, such that all the non-perturbative effects observed in the simulations arise only from the Auger mechanism and other electronic processes.  

Time-dependent simulations can resolve the spectrum as a function of the pulse duration to study the transient dynamics. Our calculations show departures from the two-step and one-step predictions, particularly in the regime of small Auger coupling $\Gamma$ and large $U_c$ and $t_{probe}$, corresponding to a situation where the screening of the core-hole takes a much longer time. We observe evidence that the core-hole attracts a polarization cloud of doublons that are responsible for a shift of spectral weight to high energies, as also observed in pump-probe simulations\cite{Rausch2019}. 

Our results indicate that in order to apply the two-step formula it is crucial to include the core-hole potential and to account for contributions originating from particle-hole excitations. One can therefore interpret the spectrum as a superposition of the two-step spectrum at a fixed density $n$ and a higher one $n+\delta$. This rule of thumb may be of help aiding experiments and their interpretation.

\bmhead{Acknowledgments}
We thank M. Eckstein, G. Sawatzky, B. Barbiellini and A. de la Torre for illuminating discussions. We acknowledge generous computational resources provided by Northeastern University's Discovery Cluster at the Massachusetts Green High Performance Computing Center (MGHPCC). AN acknowledges the support from Compute Canada and the Advanced Research Computing at the University of British Columbia, where part of simulations were performed. AN is supported by the Canada First Research Excellence Fund. AEF acknowledges the U.S. Department of Energy, Office of Basic Energy Sciences for support under grant No. DE-SC0014407. 

\begin{appendices}
\section{Derivation of the one-step formula} \label{appendix}

Following the seminal work of Gunnarsson and Sch\"{o}nhammer\cite{Gunnarsson1980}, the one-step 
formula for the Auger electron emission can we written as
\begin{align}
I^{\sigma}_{Auger} &= -\frac{1}{\pi}\sum_{\textbf{p}}|\tau_{\textbf{p},0}|^2 \textrm{Im}\Big[\langle\phi_{\textbf{p}}|A^\dag_{\textbf{k}\sigma,0}\times\nonumber\\ 
&\frac{1}{\omega_{\text{in}}-\epsilon_{\textbf{k}}-\epsilon_{\textbf{p}}-H_0+E_0+i\eta}A_{\textbf{k}\sigma,0} |\phi_{\textbf{p}}\rangle\Big],
\end{align}
where we have defined (neglecting all interatomic processes) 
$A_{\textbf{k}\sigma,0} = V_{\textbf{k}\sigma,0} \Delta_0 d^\dag_0$ as the Auger operator with matrix element 
\begin{align}
V_{\textbf{k}\sigma,0}&=\frac{e^2}{4\pi\epsilon_0}\int_{\textbf{r}_1}\int_{\textbf{r}_2} \Psi^{*}_{\uparrow}(\textbf{r}_2-\textbf{R}_0) \Psi^{*}_{\downarrow}(\textbf{r}_1-\textbf{R}_0)\frac{1}{|\textbf{r}_1-\textbf{r}_2|}\times\nonumber\\ 
\times &\Phi_{\textbf{k}\sigma}(\textbf{r}_1)\psi_{\sigma,c}(\textbf{r}_2-\textbf{R}_0)  
\end{align}
between a scattering state $\Phi_{\textbf{k}\sigma}(\textbf{r})$ with momentum $\textbf{k}$ and energy $\epsilon_{\textbf{k}}$, the core-level wave-function $\psi_{\sigma,c}(\textbf{r}-\textbf{R}_0)$, and the 
valence band wave functions $\Psi_{\sigma}(\textbf{r}-\textbf{R}_0)$.
Above, we also have defined the state 
\begin{equation}
 |\phi_{\textbf{p}}\rangle = \frac{1}{\epsilon_{\textbf{p}}-\omega_{\text{in}}-H_0-H_c+E_0+i\eta}d_0 |N,0\rangle  
\end{equation}
coming from the first phase of the Auger process. To derive Eqns~\ref{eq:1stepb}-\ref{eq:1stepc} in the main text, we
have defined $\epsilon_{\textbf{p}}\equiv\omega_{\text{photo}}$ and $\epsilon_{\textbf{k}}\equiv\omega_{\text{photo}}-\omega_{\text{Auger}}-E_c$.

Assuming that the first photoelectron has been
emitted at the Fermi edge (the sum over intermediate continuum states $\textbf{p}$ can been simplified with a single contribution at the emission \emph{edge}), we evaluate with dDMRG the following quantity
\begin{align}
I_{Auger}(\omega)&=-\frac{1}{\pi}\textrm{Im}\Big[\langle\phi_{\textrm{edge}}|\Delta^\dag_{0}\frac{1}{\omega-H_0+E_0+i\eta}\Delta_{0} |\phi_{\textrm{edge}}\rangle\Big],
\end{align}
where we have defined the quantity $\omega\equiv\tilde{\omega}_{\textbf{pk}}=
\omega_{\text{in}}-\epsilon_{\textbf{p}}-\epsilon_{\textbf{k}}$
and eliminated the core-level operators inside the vector $|\phi_{\textrm{edge}}\rangle$
\begin{equation}
 |\phi_{\textrm{edge}}\rangle = \frac{1}{\omega_{\text{edge}}-H_0-H_c+E_0+i\eta} |N,0\rangle.
\end{equation}
\end{appendices}



\end{document}